\documentstyle[12pt]{article}

\begin{document}

\begin{center}
\large
{\bf Statistical properties of high-lying chaotic eigenstates}\\
\vspace{0.3in}
\normalsize
Baowen Li\footnote{e-mail Baowen.Li@UNI-MB.SI}
 and Marko Robnik\footnote{e-mail Robnik@UNI-MB.SI}\\

\vspace{0.2in}
Center for Applied Mathematics and Theoretical Physics,\\
University of Maribor, Krekova 2, SLO-62000 Maribor, Slovenia\\
\end{center}

\section{Introduction}

The field of quantum chaos is developing very fast and there has been
substantial progress in our understanding of generic properties of
eigenstates in classically nonintegrable and chaotic bound systems.
In contrast to the theoretical description of the energy spectra
(and of other quantal observables) where we have now a rather complete
understanding  of the spectral statistical universality classes and
also of statistics in the transition region between integrability and
ergodicity (i.e. going from Poisson to GOE/GUE), we are still far from
a correspondingly complete knowledge of generic and statistical
properties of the wavefunctions. For a few recent reviews see contributions
in  Giannoni {\em et al} (1991), Gutzwiller's book (1990), the papers
in Casati {\em et al} (1993) and also the review on statistical
properties of energy spectra by Robnik (1994).
\\\\
In order to understand the wavefunctions especially in the semiclassical
limit it is intuitively very appealing to use the so-called {\em
Principle of Uniform Semiclassical Condensation} (PUSC) of the Wigner functions
(of the eigenstates) which is implicit in (Berry 1977a): As $\hbar\rightarrow
 0$ we assume that the Wigner function of a given eigenstate uniformly
(ergodically) condenses on the classical invariant object on which the
classical motion is ergodic and which supports the underlying quantal state.
Such an object can be e.g. an invariant torus, a chaotic  region as a
proper subset of the energy surface, or the entire energy surface if the
system has ergodic dynamics there.
\\\\
In classically integrable systems the eigenfunctions possess a lot of
ordered structure {\em globally} and {\em locally}: Applying PUSC the
average probability density in the configuration space is seen to be
determined by the projection of the corresponding quantized invariant
torus onto the configuration space, which implies the global order.
Moreover, the local structure is implied by the fact that the wavefunction
in the semiclassical limit is locally a superposition of a finite number
of plane waves (with the same wavenumber as determined by the classical
momentum).
\\\\
In the opposite extreme of a classically ergodic system PUSC predicts
that the average probability density is determined by the microcanonical
Wigner function. Its local structure is spanned by the superposition of
infinitely many plane waves with random phases and  equal wavenumber.
The random phases might be justified by the classical ergodicity and
this assumption, originally due to Berry (1977b), is a good starting
approximation which immediately predicts locally the Gaussian randomness
for the probability amplitude distribution. Berry (1977b) has also
calculated the autocorrelation function of semiclassical chaotic (ergodic)
wavefunctions which we will discuss later on in detail. One major surprise
in this research was Heller's discovery (1984) of scars of unstable classical
periodic orbits in classically ergodic systems. The scar phenomenon
is of course a consequence of subtle correlations in the quantal phases.
This has been analyzed and discussed by Bogomolny (1988) and Berry (1989)
in the context of the Gutzwiller periodic orbit theory. The insufficiency
of the single-periodic-orbit theory of scars has been discussed by Prosen
and Robnik (1993a) in a study of the transition region between integrability
and
chaos.
\\\\
In the generic case of a KAM-like system with mixed classical dynamics
the application of PUSC is again very useful and has a great predictive
power. Here the states can be classified as either regular (they "live"
on a quantized invariant torus) or irregular (they "live" on a chaotic
invariant region), quite in agreement with Percival's (1973) speculative
prediction, which has been recently carefully re-analyzed by Prosen and
Robnik (1994a). In this case PUSC implies asymptotic ($\hbar\rightarrow 0$)
statistical independence of level series (subsequences) associated with
different regular and irregular components. This picture has been used
by Berry and Robnik (1984) to deduce the resulting energy level statistics
in such generic Hamilton systems with mixed  classical dynamics, especially
the level spacing distribution.
In the recent work Prosen and Robnik (1994b) have numerically confirmed
the applicability of the Berry-Robnik theory and also explained the Brody-like
behaviour (as discovered and described in (Prosen and Robnik 1993b)) before
reaching the far semiclassical limit.

\section{The definition of the billiard system and the numerical technique}

In the present paper we study the chaotic wavefunctions in the 2-dim billiard
system whose domain ${\cal B}$ (in w-plane) is defined by the complex quadratic
conformal map of the unit disk (in z-plane), namely
\begin{equation}
{\cal B}_{\lambda} = \{w| w = z + \lambda z^2,\quad |z| \le 1\}.
\label{eq:cm}
\end{equation}
as introduced by Robnik (1983,1984) and further studied by Prosen and Robnik
(1993b). See also (Hayli {\em et al} 1987) and (Bruus and Stone 1994,
Stone and Bruus 1993, 1994). Following most people in the field we shall refer
to it as Robnik billiard.
As the shape parameter $\lambda$ changes from $0$ to $1/2$ this system
goes from the integrable case of the circular billiard continuously
through a KAM-like regime to an almost ergodic regime at large $\lambda$.
At $\lambda \le 1/4$ the boundary is convex and therefore the Lazutkin like
caustics and invariant tori (of boundary glancing orbits) exist.
At $\lambda \ge 1/4$ the billiard was speculated (based on numerical
evidence in (Robnik 1983)) to become ergodic, which has been disproved
by Hayli {\em et al} (1987): Close to $\lambda\ge 1/4$ there are still some
stable periodic orbits surrounded by very tiny stability islands.
On the other hand, for $\lambda=1/2$ (the cardiod billiard) the ergodicity
and mixing have been rigorously proved by Markarian (1993). Nevertheless,
at large values of $\lambda$, say $\lambda=0.375$ (which we study
exclusively in the present paper)
the numerical evidence does not exclude the possibility of ergodicity:
If there are some tiny regions of stability, then they must be so small
that they cannot be detected at large scales.
\\\\
We want to calculate and analyze the high-lying states far
in the semiclassical limit, as high as 100,000th  eigenfunction (of even parity
which is about 200,000th when counting all states) and above,
in the regime where the classical dynamics is almost completely ergodic
(within the numerical resolution of the Poincar\'e Surface Of Section).
As mentioned above, the latter condition is satisfied at $\lambda=0.375$.
However in order to
reach the said high-lying eigenstates using the available supercomputer
facilities we had to abandon the conformal mapping diagonalization technique
developed in (Robnik 1984) and further employed by Prosen and Robnik (1993b).
Instead we have implemented the Heller's method of the plane wave decomposition
of the wavefunctions (e.g. see Heller 1991). Heller's method enables one
to go very high in the semiclassical limit (high energies) where we can then
calculate a few consecutive levels, whereas the diagonalization method
(with the conformal mapping technique) has the advantage of yielding
many levels from the ground state upwards. So, if one is interested in
significant statistical analysis the latter method is superior, whilst
when studying the individual high-lying eigenstates  the former method
is the better one.
\\\\
Let us spend just a few words on the technical aspects of this difficult
task, since to the best of our knowledge many crucial ingredients
have not been discussed in the literature so far.
To solve the Schr\"odinger equation with Dirichlet boundary condition,
\begin{equation}
\Delta \Psi + E \Psi = 0, \qquad \Psi = 0 \quad {\rm at\quad the\quad
boundary},
\label{eq:Sch}
\end{equation}
we use the superposition of plane waves with the wavevectors of the same
magnitude $k$
but with different directions. The wavefunction we used for the even parity
is
\begin{equation}
\Psi(u,v)=\sum_{j=1}^{N}a_{j}\cos(k_{ju}u+\phi_{j})\cos(k_{jv}v) ,
\label{eq:wf}
\end{equation}
where $k_{ju}=k \cos(\theta_{j})$, \quad $k_{jv}=k \sin(\theta_{j})$,
\quad $k^{2}=E$ the eigenenergy, $N$ the number of
plane waves and ${\phi}_{j}$ are {\em random phases} drawn from the interval
$[0,2\pi)$, assuming uniform distribution, and  $\theta_j=2j\pi/N$
(i.e. the direction angles of the wavevectors are chosen equidistantly).
The {\em ansatz} (\ref{eq:wf})  solves the Schr\"odinger equation
(\ref{eq:Sch}) in the {\em interior} of the billiard region, so that
we have only to satisfy the Dirichlet boundary condition.
Taking the random phases, as we discovered, is equivalent
to spreading the origins of plane waves all over the billiard region,
and at the same time this results in reducing the CPU-time by
almost a factor of ten.
For a given $k$ we put the wavefunction equal to zero at
a finite number $M$ of boundary points (primary nodes) and
equal to 1 at an arbitrarily chosen interior point. Of course, $M\ge N$.
This gives an inhomogeneous set of equations which can be solved
by matrix inversion.  Usually the matrix is very singular,  thus the
{\em Singular Value Decomposition} (SVD) method has been invoked
(Press {\em et al} 1986).  After obtaining the coefficients $a_{j}$ we
calculate the wavefunctions at other boundary points (secondary nodes).
The sum of the squares of the wavefunction at all the secondary nodes
(Heller called this sum "tension") would  be ideally zero if $k^2$ is an
eigenvalue. In practice it is a positive number. Therefore the eigenvalue
problem now is to find the minimum of the "tension". In our numerical
procedure we have looked for the zeros of
the first derivative of the tension; namely the derivative is available
analytically/explicitely from (\ref{eq:wf}) once the amplitudes $a_j$
have been found. In fact, since the SVD-method is based on finding
the least square solution of the linear equations, we can choose
$M$ larger than $N$ without running into the overdetermination problem.
This has been done indeed, with a typical choice $M=5N/3$.
It must be pointed out that the wavefunctions obtained in
this way are not (yet) normalized, due to the arbitrary choice of the
interior point where the value of the wavefunction has been arbitrarilly
set equal to unity. We therefore explicitly normalize these wavefunctions
before embarking to the analysis of  their statistical properties.
\\\\
The accuracy of this method of course depends on the number of plane waves
($N$) and on the  number of the primary nodes ($M$), and we have a considerable
freedom in choosing $N$ and $M\ge N$.
In order to reach a sufficient accuracy the experience shows that
we should take at least $N=3{\cal L}/\lambda_{de Broglie}$, and $M=5N/3$, where
${\cal L}$ is the perimeter of the billiard and $\lambda_{de Broglie}$
is the de Broglie wavelength $=2\pi/k$. With this choice we reach the double
precision accuracy (sixteen digits) for all levels of integrable
systems like rectangular billiard (where the eigenenergies can be
given trivially analytically) and the circular billiard, but also
for the billiard  ${\cal B}_{\lambda}$ for $\lambda \le 0.2$.
Also, the same choice
enabled us to calculate the 100,000th even parity eigenstate and a
few nearby eigenstates for our billiard at $\lambda=0.375$ within
an accuracy of $1\%$ of the mean level spacing (seven valid digits).
These accuracy checks were based on very careful
selfconsistent checks of the method and also on comparison of
the eigenvalues with those obtained by using
Robnik's diagnalization method.
\\\\
The advantage of this method is that, at one side, it is very flexible to
calculate the eigenvalues, and on the other hand, it is self-checkable:
The accuracy and the reliability can be checked by changing
the interior point and by changing $N$ as well as $M$.
The drawback of the method is that with
unlucky choice of the interior point and unlucky energy step size
some eigenstates may be  --- and typically are! --- missed, so that the
calculation must be repeated by using different interior points to
finally collect all the levels. The Weyl formula (with perimeter and curvature
corrections) can be used to detect the missing of levels (c.f. Bohigas 1991).
A similar numerical experience has been reported in (Frisk 1990).

\section{The wavefunctions and the probability amplitude distribution}

All the wavefunctions that we calculated and discuss here are the
even parity eigenstates of the billiard
${\cal B}_{\lambda}$ at $\lambda=0.375$.
In figure 1 we plot the even parity eigenfunction of energy $E=625084.5$,
which is about 100,010th eigenstate of even parity, as estimated by
using the Weyl formula (with perimeter and curvature corrections),
\begin{equation}
N_{\rm even}(E) = \frac{1 + 2\lambda^2}{8} E -
\frac{(1 + 2\lambda){\rm E}(\sqrt{8\lambda}/(1+2\lambda))-1}{2\pi} \sqrt{E}
- \frac{1}{24}
\label{eq:Weven}
\end{equation}
where ${\rm E}(x)$ is the complete elliptic integral of second kind
(c.f. Prosen and Robnik 1993b).

This is a good example of a chaotic quantum eigenstate, which
exhibits the characteristic filamentary structure as noticed already
by Heller {\em et al} (1987,1991), which is a consequence of the fact
that in the {\em ansatz} (\ref{eq:wf}) all plane waves (with random phases)
have the same magnitude $k$ of the
wavevector. Also, as judged by the naked eye, the average probability
density is constant only if the local averaging region is sufficiently large
in units of de Broglie wavelength: Probably we need the typical size
of at least several ten wavelengths. The local and global average
value $<\Psi^2>$ is theoretically expected to be equal to $1/{\cal A}$,
where ${\cal A}= \pi(1+2\lambda^2)$ is the area of ${\cal B}_{\lambda}$.
This is a direct consequence of the microcanonically uniform Wigner
function for this eigenstate (Berry 1977, Voros 1979, Shnirelman 1979,
see also Berry 1983), which in turn is a consequence of PUSC as explained
and discussed in the introduction. Indeed, the theoretical value of $<\Psi^2>$
is 0.24844, whereas the numerical evaluation yields 0.24832 (after
averaging over 1,145,294 grid points distributed uniformly inside
the interior of the billiard region), which can be considered as an
excellent agreement. In table 1 we compare the theoretical values and
the numerical estimates for a number of eigenstates to show that quite
generally this agreement is very good. There we give the numerical
estimate for all the lowest four moments of $\Psi$-distribution,  namely
the average $m_1=<\Psi>$, the variance $m_2= <(\Psi-m_1)^2>$, the skewness
$m_3=<(\Psi-m_1)^3>/m_2^{3/2}$ and the kurtosis $m_4=<(\Psi-m_1)^4>/m_2^2-3$.
These experimental values are compared with the theoretical values of
the Gaussian random model (see introduction) which predicts
\begin{equation}
P(\Psi) = \frac{1}{\sqrt{2\pi}\sigma} \exp\{-\frac{\Psi^2}{2\sigma^2}\},
\label{eq:Gaussian}
\end{equation}
where again according to PUSC $\sigma^2=1/{\cal A}=1/(\pi(1+2\lambda^2))$,
and ideally it should be equal to $m_2$. In figure 2(a-b) we plot
the numerical histogram for $P(\Psi)$ (merely for illustrative purposes),
and --- more importantly - the cumulative distribution
$I(\Psi)= \int_{-\infty}^{\Psi} P(x)dx$ which is compared with the
theoretical model (\ref{eq:Gaussian}). The agreement is seen to be
excellent, even in figure 2(b) where, as we shall see and discuss in
section 5, in the probability density plot of the underlying wavefunction
a scar-like feature is observed.

In addition, we have estimated the significance levels of the cumulative
distribution $I(\Psi)$ according to the Kolmogorov-Smirnov test (Press
{\em et al} 1986 p472) with
respect to the Gaussian distribution for all eigenstates listed
in table 1. It is found to be exactly 1 within five digits.
This shows again that indeed our results agree excellently with the
theoretical prediction.

Similar study of chaotic eigenfunctions has been published in Chirikov {\em
et al} (1989), where even the differences between the numerics and Gaussian
random model due to the finite dimensionality of the system have been seen.

Our results are comparable to the findings of Aurich and Steiner (1993)
who studied the chaotic wavefunctions of the quantum system whose
classical counterpart is the geodesic motion on a compact surface of
constant negative curvature, although with our numerical
wavefunctions we are considerably farther in the semiclassical limit.
So far we have not found any examples of scars in these high-lying
states around 100,000th. (However, see section 5.)
The conclusion is that scars are difficult to find since they
"live" on smaller and smaller support as $\hbar \rightarrow 0$,
or $E\rightarrow \infty$, and consequently asymptotically no longer
influence the $P(\Psi)$ distribution, as further explained in section 5.

\section{The autocorrelation function of the wavefunctions}

The mean statistical properties of chaotic wavefunctions have been
discussed, analyzed and described in the previous section. However
the question of space-correlations of a wavefunction is far from
trivial.  The correlations exist on different scales and their strength
can vary substantially. For example, in figure 1 we clearly see that there is
some kind of clustering on the scale of a few ten de Broglie wavelengths:
there are regions of this size with enhanced probability density, and
there are also regions of this size with notably depleted probability
density (holes). Not every state is like that and in figure 3 we show
another chaotic even parity eigenstate with energy $E=625,118.4$,
approximate number $N_{ev}=100,015$, again for $\lambda=0.375$.
Here the above mentioned clustering is much less pronounced and
this property is well captured in the autocorrelation function of
the eigenstates as we shall see in a moment.
\\\\
The definition of the autocorrelation function of the probability
amplitude of a given eigenstate is (Berry 1977b, 1983)
\begin{equation}
C({\bf X; q})=<\Psi({\bf q + X}/2)\Psi^{*}({\bf q - X}/2)>/
<\left|\Psi({\bf q})\right|^{2}>
\label{eq:Berry}
\end{equation}
where the local average denoted by $<\cdots>$ is taken over sufficiently
large region around ${\bf q}$ whose size is typically many de Broglie
wavelengths but still small compared with the geometrical size.
In our case the wavefunction is of course real, i.e. $\Psi^{*}=\Psi$.
It should be noted that the nominator in (\ref{eq:Berry}) is actually
the Fourier transform of the Wigner function $W({\bf q,p})$,
\begin{equation}
W({\bf q,p}) = \frac{1}{(2\pi)^{2}} \int d^{2}{\bf X}
\exp(-i{\bf p}\cdot{\bf X}) \Psi({\bf q - X}/2)\Psi({\bf q + X}/2)
\label{eq:Wigner}
\end{equation}
where we have specialized to our real $\Psi$ case, and also two degrees
of freedom and $\hbar=1$. Now using the PUSC for a chaotic state
following Berry (1977b) we assume that the Wigner function of such a
classically ergodic state is microcanonical, i.e.
\begin{equation}
W({\bf q,p}) = \frac{\delta (E-H({\bf q,p}))}
{\int d^2 q d^2 p \delta(E-H({\bf q,p}))}
\label{eq:Wignererg}
\end{equation}
Thus substituting (\ref{eq:Wignererg}) into the inversion of
(\ref{eq:Wigner}) and then into (\ref{eq:Berry})
we immediately obtain the special case of Berry's (1977b) result,
namely
\begin{equation}
C({\bf X; q})= J_{0}(ks)
\label{eq:Bessel}
\end{equation}
where $J_{0}$ is the Bessel function of zero order, $k^2$ is the eigenenergy
and $s$ is the length of ${\bf X}$. So the autocorrelation function
is isotropic, and we are going to check numerically the validity of this
theoretical prediction.
\\\\
First we would like to check the isotropy of the autocorrelation function.
To this end we have evaluated (\ref{eq:Berry}) by taking the local
average on a small strip of $20\times100$ wavelengths situated at
the center of the billiard as far as possible from the boundaries.
The results for the wavefunctions
of figure 1 and figure 3 are shown in figures 4 and 5 correspondingly.
Because of the inversion symmetry of the autocorrelation function
w.r.t. ${\bf X}$ and the reflection
symmetry of the wavefunctions $\Psi$ w.r.t. $v$ we can restrict ourselves
to the angles within the interval $[0,\pi/2]$, and we have chosen
the values between $0$ and $\pi/2$ in equal steps of $\pi/12$,
as indicated in the upper right corner of the figures.
The autocorrelation function is obviously strongly direction dependent (please
notice that the statistical noise is practically zero) and in the case of
more uniformly chaotic wavefunction of figure 3 agrees better with the
theoretical prediction (\ref{eq:Bessel}) than for the less chaotic
eigenstate of figure 1. We believe that the semiclassical periodic orbit
theory (see e.g. Casati {\em et al} 1993, T\'el and Ott 1993 and references
therein)
could explain the deviations from the  isotropy.  Our results agree
qualitatively with (Aurich and Steiner 1993) although we are
considerably higher in the semiclassical limit (by a factor of 10 or so),
and also with somewhat old results in (McDonald and Kaufman 1988, Shapiro and
Goelman 1984, Shapiro {\em et al} 1988).
\\\\
It is interesting that after averaging over many directions we get
a considerable agreement with (\ref{eq:Bessel}).
This is shown in figure 6(a-d) where we vary the size of the averaging disk
and also the number of the directions over which the average is taken.
These plots are for the eigenfunction shown in figure 1.
Both effects are clearly visible, namely the increasingly better agreement
with (\ref{eq:Bessel}) as we increase the radius of the averaging disk and/or
as we increase the number of directions. The same aspects are shown in figure
7(a-d) for the more uniformly chaotic state of figure 3. By comparing the
figures 6(a) and 7(a) we see that in the latter plot the agreement with theory
is better.

\section{Scar-like features in wavefunctions}

As we know since Heller's (1984) discovery of scars (of unstable classical
periodic orbits) in chaotic quantum eigenfunctions of classically ergodic
systems we do expect such scars to exist in all chaotic systems, but
according to the single-periodic-orbit theory (Bogomolny 1988, Berry 1989)
the scar supporting region should shrink as $\sqrt{\hbar}$ as
$\hbar\rightarrow 0$, whilst the probability density contrast remains
fixed since it is predicted to be $\hbar$-independent (Heller 1984). The
many-orbits theory (Robnik 1989) would speculatively predict the linear
scaling of the scar area with $\hbar$ as a  consequence of the interference
effects. Some phenomenological material on this topic has been recently
published in (Prosen and Robnik 1993a).
\\\\
In this short section we would like to draw attention to an interesting
scar-like feature seen in figure 3: Near the boundary, about ten de Broglie
wavelengths away, there is a thin scar-like  feature which has no simple
explanation, because due to the nonconvexity of the billiard boundary
there are no Lazutkin-like caustics and invariant tori and also no such
glancing periodic orbits. The only classical object that might be relevant for
this feature is possibly the glancing orbit which survives many bounces
while going round the boundary until reaching the non-convexity region
and flying away, becoming completely chaotic afterwards. We have
observed a few similar features in quite a few other eigenfunctions, but
we cannot offer any definite theoretical explanation so far. However, the
formalism offfered and discussed in (M\"uller {\em et al} 1993) might just
be right to quantitatively describe the role of such orbits which are
recurrent in configuration space but not periodic.

\section{Discussion and conclusions}

In this paper we have numerically calculated the high-lying chaotic states
in the Robnik billiard as high as 200,000th eigenstate and investigated
their semiclassical morphology and their statistical properties. To achieve
this we have implemented and adapted Heller's method of plane wave
decomposition which has been further developed and its accuracy carefully
checked. Similar to other workers (e.g. Aurich and Steiner 1993) we reach
the following conclusions. In such high-lying eigenstates the scars are
hardly detectable since so far we have not found any of them. The average
probability density is globally in excellent agreement with the theoretical
semiclassical (and classical!) prediction. The Gaussian
random model for the local statistical properties of the wavefunctions is
generally excellent, in spite of the characteristic filamentary structure
and the relevant clustering of probability density on the scales of  a few
ten de Broglie wavelengths. This has been found by comparing the theoretical
and the numerical distributions and also by the comparison of the lowest
four moments and the evaluation of the Kolmogorov-Smirnov test. The
autocorrelation function nicely captures the clustering property, is found
to be strongly direction dependent in contradistinction with Berry's
(1977b) isotropic prediction,  but after averaging over many directions
the agreement with Berry's theory is recovered. Finally we should mention
that in some of the eigenstates we discovered scar-like features resembling
the whispering gallery modes for which we do not have a proper theoretical
explanation.
\\\\
Our current and future work deals with the systematic search for the scars
and the analysis of their geometry and scaling properties with $\hbar$.
On theoretical side the present paper stimulates further work on the scar
theory for which we expect improvement when many-orbits theory will
be set up following  the suggestions in (Robnik 1989, Prosen and Robnik 1993a).
Moreover, we believe  that the application of the Gutzwiller's (one-)
periodic orbit theory could explain in detail the anisotropies of the
autocorrelation function. Our work also shows that there is still
much interesting structure in the range of a few ten de Broglie wavelengths
in chaotic wavefunctions which calls for a more refined statistical
description.

\section*{Acknowledgments}

We wish to thank Boris V. Chirikov for helpful comments and relevant
references. The financial support by the Ministry of Science and
Technology of the Republic of Slovenia is gratefully acknowledged.

\vfill
\newpage
\section*{References}
Aurich R and Steiner F 1993 {\em Physica D} {\bf 64}, 185\\\\
Berry M V 1977a {\em Phil. Trans. Roy. Soc. London} {\bf 287} 237\\\\
Berry M V 1977b {\em J. Phys. A: Math. Gen.} {\bf 10} 2083\\\\
Berry M V 1983 in  in {\em Chaotic Behaviour of Deterministic Systems
(Proc. NATO ASI Les Houches Summer School)} eds
Iooss G, Helleman R H G and Stora R (Amsterdam: Elsevier) p171\\\\
Berry M V 1989 Proc. Roy. Soc. London {\bf A423} 219\\\\
Berry M V and Robnik M 1984 {\em J. Phys. A: Math. Gen.} {\bf 17} 2413\\\\
Bohigas O, 1991  in {\em Chaos and Quantum Systems (Proc. NATO ASI Les Houches
Summer School)} eds M-J Giannoni, A Voros and J Zinn-Justin,
(Amsterdam: Elsevier) p87\\\\
Bogomolny E B 1988 {\em Physica D} {\bf 31} 169\\\\
Bruus H and Stone A D 1994 {\em Preprint} Dept. Phys. Yale University\\\\
Casati G, Guarneri I and Smilansky U eds. 1993
{\em Quantum Chaos} (North-Holland) \\\\
Chirikov B V, Izrailev F M and Shepelyansky D L 1988 {\em Physica} {\bf D
33} 77\\\\
Frisk H 1990 Nordita {\em Preprint}.\\\\
Giannoni M-J, Voros J and Zinn-Justin eds. 1991 {\em Chaos and Quantum Systems}
(North-Holland)\\\\
Gutzwiller M C 1990 {\em Chaos in Classical and Quantum Mechanics} (New York:
Springer)\\\\
Hayli A, Dumont T, Moulin-Ollagier J and Strelcyn J M 1987 {\em J. Phys. A:
Math. Gen} {\bf 20} 3237\\\\
Heller E J 1984 {\em J Phys. Rev. Lett.} {\bf 53} 1515\\\\
Heller E J, O'Connor P W and Gehlen J 1987
{\em Phys. Rev. Lett.} {\bf 58} 1296\\\\
Heller E J 1991  in {\em Chaos and Quantum Systems (Proc. NATO ASI Les Houches
Summer School)} eds M-J Giannoni, A Voros and J Zinn-Justin,
(Amsterdam: Elsevier) p547\\\\
Markarian R 1993 {\em Nonlinearity} {\bf 6} 819\\\\
McDonald S W and Kaufman A N 1988 {\em Phys. Rev. A} {\bf 37} 3067\\\\
M\"uller K, H\"onig A and Wintgen D 1993 {\em Phys. Rev. A} {\bf 47}
3593\\\\
Percival I C 1973 {\em J. Phys. B: At. Mol. Phys.} {\bf 6} L229 \\\\
Press  W H, Flannery B P, Teukolsky S A and Vetterling W T
1986 {\em Numerical Recipes} (Cambridge University Press)\\\\
Prosen T and Robnik M 1993a {\em J. Phys. A: Math. Gen.} {\bf 26} 5365\\\\
Prosen T and Robnik M 1993b {\em J. Phys. A: Math. Gen.} {\bf 26} 2371\\\\
Prosen T and Robnik M 1994a submitted to {\em J.Phys.A: Math. Gen.}
in April 1994\\\\
Prosen T and Robnik M 1994b submitted to J. Phys.A. Gen in March 1994\\\\
Robnik M 1983 {\em J. Phys. A: Math. Gen.} {\bf 16} 3971\\\\
Robnik M 1984 {\em J. Phys. A: Math. Gen.} {\bf 17} 1049\\\\
Robnik M 1989 {\em Preprint} Institute of Theoretical Physics, University
of California Santa Barbara\\\\
Robnik M 1994 {\em J. Phys. Soc. Japan Suppl.} {\bf 63} \\\\
Shapiro M and Goelman G 1984 {\em Phys. Rev. Lett.} {\bf 53} 1714 \\\\
Shapiro M, Ronkin J and Brumer P 1988 {\em Chem. Phys. Lett.} {\bf 148} 177\\\\
Shnirelman A L 1979 {\em Uspekhi Matem. Nauk} {\bf 29} 181\\\
Stone A D and Bruus H 1993 {\em Physica B} {\bf 189} 43\\\\
Stone A D and Bruus H 1994 {\em Surface Science} in press\\\\
T\'el T and Ott E 1993 {\em Chaos Focus Issue on "Chaotic Scattering"}
{\bf 3}\\\\
Voros A 1979 {\em Lecture Notes in Physics} {\bf 93} 326\\\\

\vfill
\newpage
\section*{Table}

\bigskip
\bigskip

\noindent {\bf Table 1:}
The average, variance, skewness and kurtosis of a few eigenstates nearby
100,000th eigenstate of even parity in comparison with theoretical values.
 $N_{ev}(E)$  is given by the Weyl formula (\ref {eq:Weven}).
The significance levels of the Kolmogorov-Smirnov test for all
eigenstates listed in this table are exactly one within 5 digits.

\bigskip
\bigskip
\bigskip

\begin{tabular}{|llrlrr|} \hline
$E$ &   $N_{ev}(E)$ &  Average & Variance & Skewness & Kurtosis\\ \hline
$625040.6$ & $100003$ & $ 0.00002$ & $0.24828$ & $-0.00211$ & $ 0.11607$\\
$625058.4$ & $100006$ & $ 0.00001$ & $0.24836$ & $ 0.00441$ & $ 0.06576$\\
$625084.5$ & $100010$ & $ 0.00001$ & $0.24832$ & $-0.00160$ & $ 0.06809$\\
$625099.5$ & $100012$ & $-0.00001$ & $0.24834$ & $-0.00179$ & $ 0.03638$\\
$625118.4$ & $100015$ & $ 0.00001$ & $0.24838$ & $ 0.00460$ & $-0.01415$\\
$625161.9$ & $100022$ & $ 0.00003$ & $0.24837$ & $ 0.00347$ & $ 0.02424$\\
$625172.8$ & $100024$ & $ 0.00007$ & $0.24839$ & $-0.00248$ & $ 0.04836$\\
$625182.1$ & $100025$ & $ 0.00000$ &$0.24854$ &$-0.00135$ & $0.03624$\\ \hline
$ Gaussian$ &   & $0.0$ &  $ 0.24844$  & $0.0$ & $0.0$  \\ \hline
\end{tabular}

\newpage
\section*{Figure captions}

\bigskip
\bigskip

\noindent {\bf Figure 1:}
The probability density plot for the even parity eigenstate
with $E=625,084.5$, for $\lambda=0.375$, and with the estimated sequential
number using the Weyl formula (\ref{eq:Weven}) equal to 100,010.
The contours are plotted at ten equally spaced steps between zero
and the maximum value. In this geometry the unit length is 126 de Broglie
wavelengths.

\bigskip
\bigskip

\noindent {\bf Figure 2:}
The probability distribution function and cumulative distribution function
of eigenstates $E=625,084.5$ (approximately 100,010th even parity state) (a),
and $E=625,118.4$ (approximately 100,015th even parity state)
(b), in comparison with
the Gaussian random model (\ref {eq:Gaussian}).
In the top diagrams we show the histograms compared with
theoretical curve (\ref{eq:Gaussian}), and in the lower diagrams we
show the cumulative amplitude distribution function $I(\Psi)$. Three
small boxed regions are displayed in the corresponding magnified windows.
Here the difference between the theoretical and the numerical curves is
hardly visible since the agreement is so good.

\bigskip
\bigskip

\noindent {\bf Figure 3:}
The probability density plot for the even parity eigenstate
with $E=625,118.4$, for $\lambda=0.375$, and the estimated sequential
number using the Weyl formula (\ref{eq:Weven}) equal to 100,015.
The contours are plotted at ten equally spaced steps between zero
and the maximum value. In this geometry the unit length is 126 de Broglie
wavelengths.

\bigskip
\bigskip

\noindent {\bf Figure 4:}
The autocorrelation function $C({\bf X},{\bf q})$ of the eigenstate
in figure 1. $C({\bf X},{\bf q})$ is plotted against $ks$ for seven
different angles of ${\bf X}$ w.r.t the abscissa; here
$s=\left|{\bf X}\right|$. The angle and the averaging
strip are indicated in the upper right corner of the figure. From (a) to (g)
the angle goes from $0$ to $\pi/2$ with the increment of $\pi/12$.
The dashed curve is the theoretical prediction (\ref{eq:Bessel}),
namely $J_0(ks)$ , whilst the full curve denotes
the numerical result. The local average has been taken on a strip of
$20\times100$ de Broglie wavelengths. For a fixed $ks$ about 150,000
grid points inside the strip have been used to calculate $C$.
The reference point ${\bf q}$ is fixed at $(0.4,0.0)$ which is probably
sufficiently far away from the billiard boundary.

\bigskip
\bigskip

\noindent {\bf Figure 5:}
The same as figure 4 but for the eigenstate of figure 3.

\bigskip
\bigskip

\noindent {\bf Figure 6:}
The autocorrelation function after averaging over many directions ${\bf X}$
of the eigenstate in figure 1. $C({\bf X},{\bf q})$ is plotted against
$ks$ for three different averaging disks and different number of directions,
where $s=\left|{\bf X}\right|$.
The averaging disk in (a), (b), (c) and (d) has the diameter
of 200, 100, 50 and 100 de Broglie wavelengths, respectively. In (a), (b) and
(c) the number of directions is 200, whilst in (d) it is 400.

\bigskip
\bigskip

\noindent {\bf Figure 7:}
The same as in figure 6 but for the eigenstate in figure 3.
\end{document}